# Observation and dynamic control of a new pathway of $H_3^+$ formation

Yonghao Mi[1,*], Enliang Wang[2,3], Zack Dube[1], Tian Wang[1], A. Yu. Naumov[1], D. M. Villeneuve[1], P. B. Corkum[1], André Staudte[1*]

[1]*Joint Attosecond Science Laboratory, National Research Council and University of Ottawa, 100 Sussex Drive, Ottawa K1A 0R6, Canada*

[2]*Hefei National Laboratory for Physical Sciences at the Microscale and Department of Modern Physics, University of Science and Technology of China, Hefei, Anhui 230026, China*

[3]*Max Planck Institut für Kernphysik, Saupfercheckweg, 1, 69117 Heidelberg, Germany*

e-mail: ymi@uottawa.ca; andre.staudte@nrc-cnrc.gc.ca

**Abstract**

We propose and experimentally demonstrate that the trihydrogen cation ($H_3^+$) can be produced via single photoionization of the molecular hydrogen dimer ($H_4$). Using near-infrared, femtosecond laser pulses and coincidence momentum imaging, we find that the dominant channel after single ionization of the dimer is the ejection of a hydrogen atom within a few hundred femtoseconds, leaving an $H_3^+$ cation behind. The formation mechanism is supported and well reproduced by an ab-initio molecular dynamics simulation. This is a new pathway of $H_3^+$ formation from ultracold hydrogen gas that may help explain the unexpected high abundance of $H_3^+$ in the interstellar medium in the universe.

**Introduction**

The trihydrogen cation ($H_3^+$), structured by three protons and two electrons, is one of the most abundant ions in the universe. As a proton donor, $H_3^+$ initiates most chemical reactions in interstellar space, making significant contributions to the formation of almost all molecules[1-3]. Following ionization of interstellar $H_2$ by cosmic rays, the subsequent ion-molecule reaction ($H_2 + H_2^+ \rightarrow H_3^+ + H$) is considered to be the main pathway of $H_3^+$ production in the interstellar medium. In 1996 and 1998, $H_3^+$ was detected in dense[4] and in diffuse interstellar molecular clouds[5]. Surprisingly, the detected $H_3^+$ column density in diffused clouds is a few orders of magnitude higher than the expected value derived from a model that considers the production and destruction rates of $H_3^+$ in those diffused clouds[1]. This is known as the $H_3^+$ problem and the enigma of the pervasive $H_3^+$ in the interstellar medium has been challenging to understand.

When first discovered by J. J. Thomson with a prototypic mass-spectrometer in 1911[6], $H_3^+$ was produced in an electrical discharge in $H_2$ gas via the ion-molecule reaction between $H_2^+$ and $H_2$ [7]. $H_3^+$ can also be produced by doubly ionizing small organic molecules using electron- or ion- impact ionization[8–11] or photoionization with intense laser fields[8, 12–21]. This formation mechanism has been attributed to $H_2$ roaming and the time scale of this process has been investigated[11, 13, 18–21]. Recently, the formation of $H_3^+$ from nanoparticle-adsorbed water molecules has been observed[22].

As the most abundant element in the universe, huge amounts of atomic hydrogen exist in our galaxy in warm ($10^3 – 10^4$ K) and cool (50 – 100 K) regions, while cold (10 – 30 K) regions are considered to be dominated by dense clouds of molecular hydrogen[23]. The presence of $H_2$ in these clouds has been attributed to the catalytic action of dust particles[24]. The same mechanism could allow the formation of $H_2$ dimers. $H_2$

dimers, with an actual binding energy of ~ 3 cm$^{-1}$, are among the most weakly bound dimers in nature. Therefore, H$_2$ dimers are elusive to spectroscopic detection, yet they have been found in the upper atmospheres of Jupiter and Saturn[25-28].

In this paper, we propose a new pathway of H$_3^+$ formation under laboratory conditions, which proceeds through single ionization of an H$_2$ dimer (H$_4$) and a subsequent proton transfer process in the dimer. Throughout this work, deuterium molecules (D$_2$) are used in calculation and experiment. This choice is made for experimental reasons: by measuring D$_3^+$ instead of H$_3^+$, we can avoid the ambiguity of the H$_3^+$ mass spectrum peak with the ever-present contaminant HD$^+$. Moreover, having the exact same electronic structure, D$_4$ only differs from H$_4$ in the reduced speed of the nuclear dynamics making it also more amenable for our pump-probe scheme.

**Results and Discussion**

We first calculated the potential energy surface (PES) of D$_4$. As shown in Fig. 1a, the coordinates R$_1$ and R$_2$ are the center-of-mass distance between the two molecules of the dimer and the internuclear distance of one of the D$_2$ molecules, respectively. The point A on the lower PES shows the stationary point (equilibrium geometry) of the ground state of D$_4$. When an electron is removed, the molecular dimer is projected onto the ground state of D$_4^+$ and moves along the pathway marked by the yellow line. The D$_4^+$ ground state has an energy minimum (point B) with a potential energy barrier of about 0.25 eV (see details in the Supplementary Note 1), which is lower than the potential energy of the Franck-Condon region. Therefore, D$_4^+$ is not a stable molecular ion and dissociates into a D$_3^+$ ion and a deuterium atom (point C).

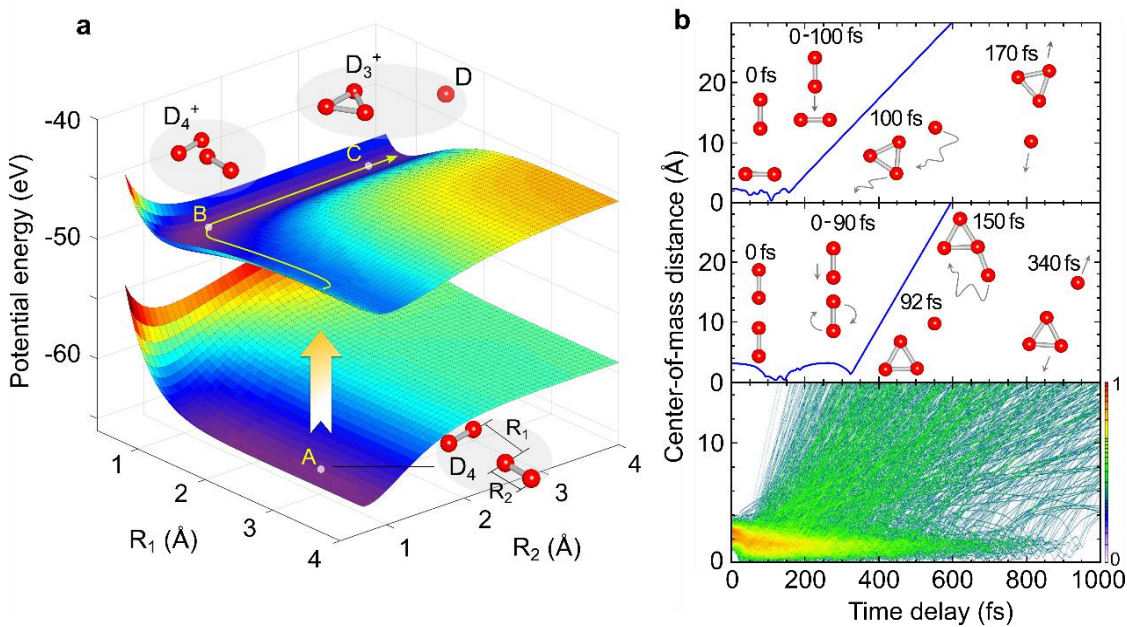

**Fig. 1 | Potential energy surface (PES) and trajectory simulations. a,** Calculated PES of a $D_2$ dimer and the schematic pathway of $D_3^+$ formation. **b,** Simulation results of the time-dependent distance between $D_3^+$ and D for a T-shaped (top panel) and a parallel-aligned (middle panel) $D_2$ dimer, and simulated trajectories (center-of-mass distance between $D_3^+$ and D) of $D_3^+$ formation for 892 $D_2$ dimer cations (bottom panel).

With the knowledge of this possible pathway, we carried out an ab-initio molecular dynamics simulation to extract the temporal information of the $D_3^+$ formation. 1000 trajectories were simulated, and 892 trajectories led to $D_3^+$ formation. Figure 1b shows the simulated center-of-mass distance between $D_3^+$ and D as a function of the time delay. In the top two panels of Figure 1b we show two typical trajectories for the T-shape and parallel-aligned $D_2$ dimer. These two trajectories start from their respective equilibrium geometries without any initial internal energy. For ionization of a T-shape dimer, the $D_2^+$ moves towards the $D_2$ and a deuteron is attracted by the $D_2$. At 100 fs, three deuterons start bonding together while all four deuterons are still close to each other. The produced $D_3^+$ and D separate after 170 fs, as indicated by the trajectory in the top panel of Fig. 1b. For a parallel-aligned dimer, the formation process is the same as that of the T-shape dimer. However, the formation time is about twice as long (see the middle panel of Fig. 1b). The complete dynamics of the $D_3^+$ formation for both geometries of the dimer can be found in the supplementary videos. The bottom panel of Fig. 1b shows the simulated time-dependent trajectories (center-of-mass distance between $D_3^+$ and D) for the 892 dimer cations leading to $D_3^+$. For comparison, the simulation was also performed on $H_2$ dimers and 83.1% of the $H_2$ dimers produce $H_3^+$. The result of the simulation on $H_2$ dimers is shown in the Supplementary Note 4.

Next, we turn to the experiment for the formation of trihydrogen cations from hydrogen molecular dimers using $D_2$. The experiment was performed in a cold-target recoil-ion momentum spectroscopy (COLTRIMS) reaction microscope[29]. The $D_2$ dimers were prepared in a cold molecular beam via supersonic expansion of $D_2$ gas into a high-vacuum ($10^{-11}$ mbar) chamber through a 10-μm nozzle and a skimmer (see Fig. 2a). Details of the experimental setup can be found in the Methods section.

The time-of-flight (TOF) spectrum of the photoions was measured. As shown in Fig. 2b, the dominant peak is $D_2^+$ (TOF ~ 2250 ns) which is produced by single ionization of the $D_2$ molecules in the gas jet. The multiple peaks centered at around 1590 ns in the spectrum correspond to $D^+$ ejected during dissociation of $D_2^+$. The sharp peak at 1950 ns is $HD^+$ originating from the HD in the $D_2$ gas cylinder. Centered at around 2750 ns, a broad peak with a mass-to-charge ratio of 6 was observed. This is the first evidence of the observation of $D_3^+$ produced from $D_4$.

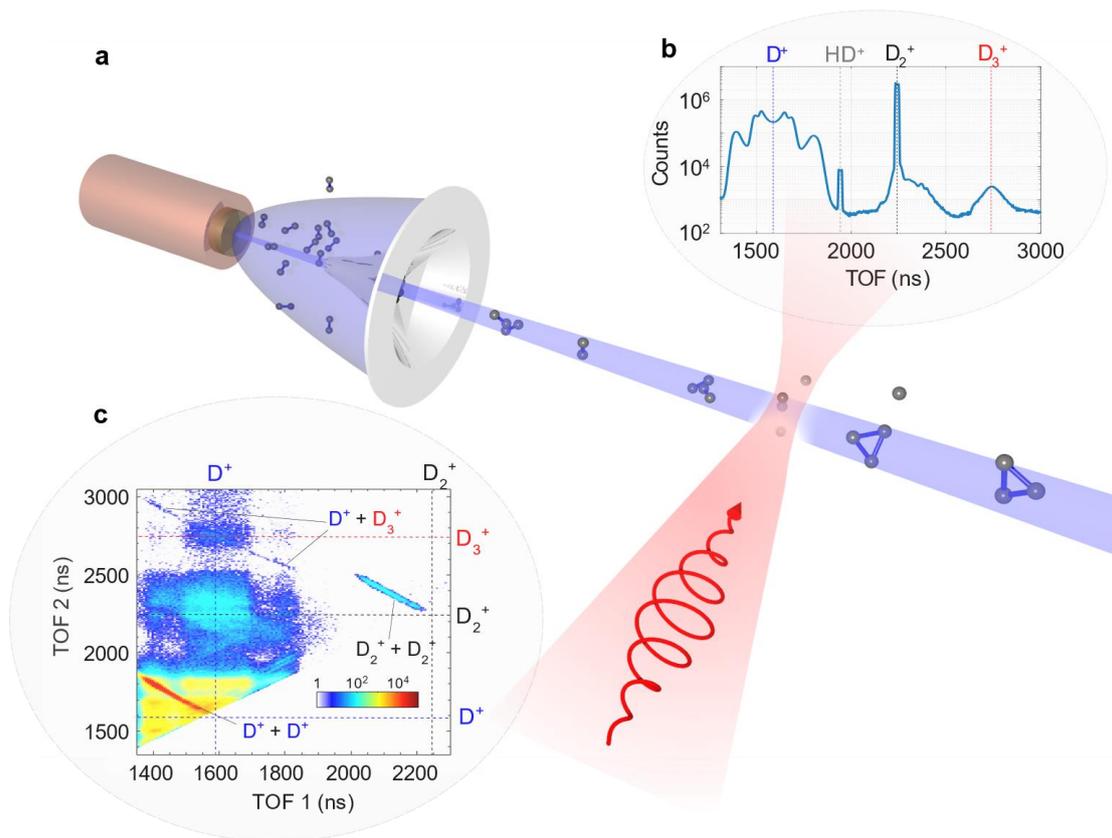

**Fig. 2 | Sketch of the experiment and TOF spectra. a,** The $D_2$ dimers are generated in the supersonic expansion of $D_2$ gas through a pre-cooled (60 K) nozzle and a skimmer. The dimers are ionized by 25 fs, 790 nm, circularly polarized laser pulses. **b,** Measured time-of-flight (TOF) spectrum of the photoions. **c,** Measured photoion-photoion coincidence (PiPiCo) spectrum.

To investigate the break-up mechanism of $D_4$, the photoion-photoion coincidence (PiPiCo) spectrum was recorded (see Fig. 2c). The sharp diagonal line on this figure indicates that the first and the second ion originate from a two-body fragmentation of a molecule. The center TOFs of $D^+$, $D_2^+$ and $D_3^+$ are marked with dashed lines for better characterization of the coincidence between these photoions. As shown in Fig. 2c, besides the $D^+ + D^+$ correlation which indicates the Coulomb explosion of $D_2$ molecules, we also observed $D_2^+ + D_2^+$ and $D^+ + D_3^+$ correlations. These two channels originate from the double ionization and subsequent break-up of $D_2$ dimers. The correlation on the PiPiCo spectrum confirms the new pathway of $D_3^+$ formation. We note that the probability of double ionization is much smaller than that of single ionization. Therefore, the yield of $D^+ + D_3^+$ is much lower compared to the yield of $D_3^+$ that is produced from single ionization of $D_4$ ($D_4^+ \rightarrow D_3^+ + D$). In this experiment, $D_4^+$ was not observed in the TOF spectrum. This agrees well with the simulation results: $D_4^+$ is not stable and will dissociate into $D_3^+$ and D.

Is it possible to measure the time scale of the $D_3^+$ formation process? To answer this question, we designed a pump-probe experiment in which the circularly polarized laser pulse ($3 \times 10^{14}$ W/cm$^2$) serves as the pump pulse, and a delayed, weak pulse ($5 \times 10^{13}$ W/cm$^2$) with linear polarization was introduced as the probe. The circularly polarized pump pulse triggers single ionization of $D_4$ while minimizing alignment-selective ionization. Furthermore, circular polarization suppresses recollision-induced processes such as double ionization. The linearly polarized probe pulse is too weak to ionize $D_4$, however, it can work as a disruptive pulse[20] which perturbs the intermediate $D_4^+$ cation either by dissociating or by secondary ionization and thereby probes the dissociation dynamics.

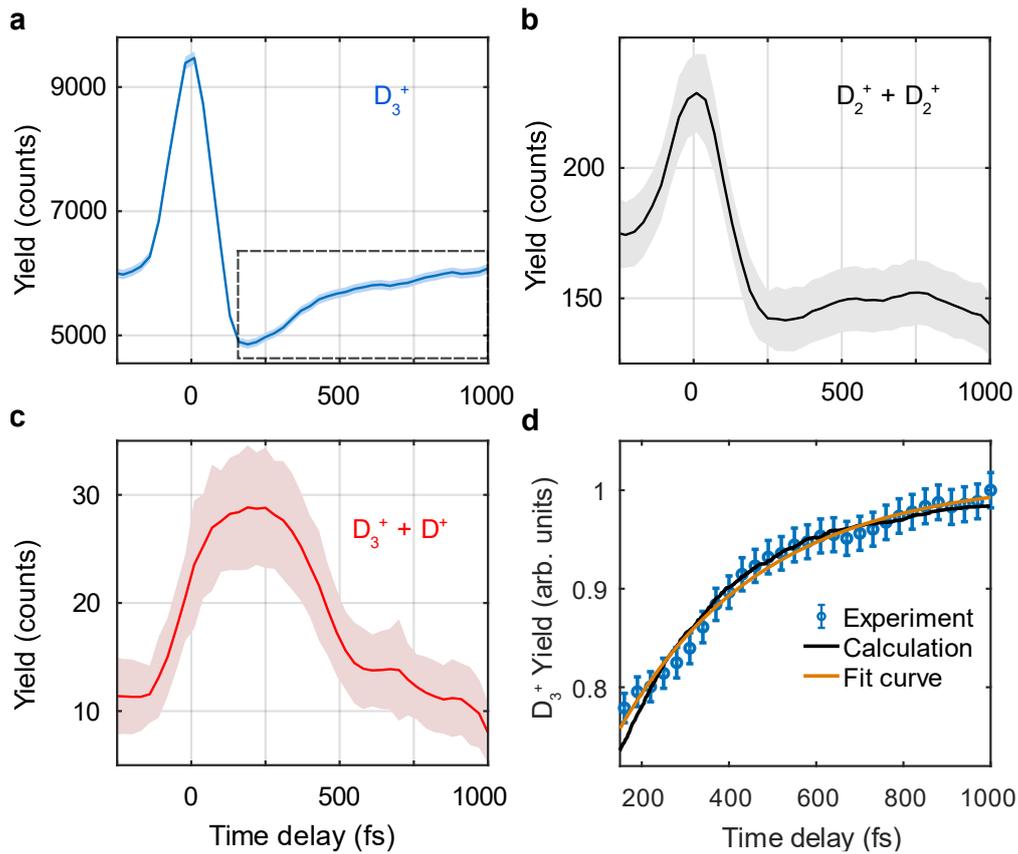

**Fig. 3 | Experimental results of the time-dependent yields for different channels.** Yield of $D_3^+$ (**a**), $D_2^+ + D_2^+$ (**b**), and $D_3^+ + D^+$ (**c**) as a function of the pump-probe time delay (shaded area given by the error bars). **d,** Experiment, calculation and fit curve of the time-dependent $D_3^+$ yield for the range between 150 and 1000 fs.

Figure 3a shows the yield of $D_3^+$ as a function of the pump-probe delay for the time-delay range between -250 and 1000 fs. The positive and negative delays indicate that the pump pulse comes before and after the probe pulse, respectively. When the delay is between -150 and 150 fs, the constructive interference between the two pulses produces a higher peak intensity. Thus, the ionization rate of $D_4$ and thereby the yield of $D_3^+$ are peaked in this time window. However, the yield of $D_3^+$ increases exponentially from 150 to 1000 fs. The exponential shape can be understood qualitatively by the following $D_3^+$ formation dynamics: When an electron is removed from a $D_4$ at time zero, the $D_4^+$ starts to dissociate. In the first hundred femtoseconds, the deuterons are still in the vicinity of each other. The weak probe pulse can destroy the rearrangement process and prevent the system from evolving to a $D_3^+$ and a D atom. However, when the $D_3^+$ has already been formed at larger delays (i.e., > 400 fs), the probe pulse is not sufficiently intense to interrupt this process and thus the $D_3^+$ yield recovers as the delay increases. When the weak probe pulse arrives first, it is not strong enough to ionize $D_4$. The $D_4$ ionized by the late-coming pump pulse leads to $D_3^+$ in any case. Therefore, no significant time-dependent yield of $D_3^+$ was observed at negative delays.

Next, we investigate the ultrafast dissociation dynamics of the double-ionization channels. We show the time-resolved yields of $D_2^+ + D_2^+$ and $D^+ + D_3^+$ in Fig. 3b and c, respectively. The $D_2^+ + D_2^+$ yield does not show a distinct time-dependent variation outside the temporal overlap (>150 fs). However, the yield of $D^+ + D_3^+$ exhibits a maximum around 200 fs, when the yield of the $D + D_3^+$ channel is minimized.

Comparing the yields of the two double ionization channels with the time-resolved $D_3^+$ yield, we find that the suppression of the $D_3^+$ yield within the first few hundred femtoseconds can be at least partially explained with an enhanced yield of $D_3^+ + D^+$. However, in view of the overall low probability for double ionization (due to the low-intensity probe pulse), a dissociation of the $D_4^+$ dimer into $D^+ + D + D_2$ via bond softening[30] is likely the dominant pathway induced by the probe pulse. This bond softening is caused by a very strong AC-Stark shift of the highly polarizable molecular ion. Details of the suppression of the $D_3^+$ yield are discussed in Supplementary Note 6.

We now compare our experimental results with the calculation. For delay times smaller than 150 fs the overlap of the pump and the probe pulses causes an enhanced yield in all channels. Hence, we concentrate our attention on the time window between 150 and 1000 fs. In Fig. 3d the measured $D_3^+$ yield (normalized to the maximum value), shown by blue circles, starts out from a minimum and then grows to asymptotically recover to the value at negative time delays. We fit the $D_3^+$ yield with the growth function $Y(t) = a(1 - e^{-t/\tau}) + Y_0$, where $a$ and $Y_0$ are the amplitude and the offset of the $D_3^+$ yield, and $\tau$ is the formation time of $D_3^+$. Using this fit function, we extract the formation time $\tau = 330 \pm 55$ fs. The theoretical $D_3^+$ yield (black line) is in good agreement with the experimental data (blue circles) and the fit curve (orange line).

**Conclusion**

In conclusion, we have demonstrated in both theory and experiment that the trihydrogen cation can be produced via single ionization of a hydrogen molecular dimer. Taking a different perspective, the dimer represents a chemical reaction in waiting, which is triggered by the arrival of the ionizing pump pulse. Upon ionization the $D_4^+$ begins the transition to $D_3^+ + D$. The probe pulse modifies the potential energy surface leading the system away from the dominant reaction pathway. This is an example of the dynamic Stark-control of a chemical reaction[31]. The pump-probe experiment demonstrated that chemical reactions on isolated bimolecular system can be controlled and steered by a weak laser pulse.

While $H_2$ dimers will not be subjected to intense laser pulses in interstellar space, the ionization dynamics can be expected to proceed very similarly for high energy photons (cosmic rays) due to the electronic simplicity of $H_2$. Although the detection of $H_2$ dimers in the interstellar medium is challenging due to their weak bond, the observation of $H_2$ dimers in the atmospheres of Jupiter and Saturn, could indicate that the contribution of hydrogen dimers to the formation of $H_3^+$ has been neglected.

# Methods

## Ab-initio molecular dynamics simulation

Ab-initio molecular dynamics simulation is a technique which allows for the simulation of molecular system and processes from first principles. The simulation was performed under the extended Lagrangian molecular dynamics scheme in which the atom-centered density matrix propagation (ADMP) approach and the density functional theory (ωb97xd/aug-cc-pVDZ) were applied[32-35]. The details are described in the Supplementary Note 2-3 and only a brief introduction is shown here. The initial molecular configuration, i.e., the geometries and the velocities of every atom of $D_2$ dimer, were calculated by the Wigner distribution which traversed all the possible region on the potential energy surface of $D_2$ dimer. We assumed a vertical ionization from the neutral to cationic state of $D_2$ dimer and the sampled geometry and velocity were used as the initial configuration of the molecular dynamics simulation of the cationic state of $D_2$ dimer. In the ADMP simulation, the fictitious electron mass was 0.1 amu, and the simulation time step was 0.5 fs. Due to the low potential energy barrier of $D_2$ dimer, the anharmonic effect in the sampling cannot be ignored. By comparing the sampled internuclear distance distribution and the ab initio potential energy curve (see Supplementary Note 3 for details), we find that the anharmonicity of the intermolecular potential leads to 7% of our trajectories to start inside of the inner classical turning point. The molecular dynamics simulation showed that these events result in the same dissociation limit and may result in a shorter formation time of $D_3^+$.

## Experimental setup

In the reaction microscope, the backing pressure of the nozzle was 3 bar, and the $D_2$ gas in the nozzle was precooled to 60 K. The molecular beam intersects with a focused near-infrared laser beam (central wavelength λ = 790 nm, pulse duration τ = 25 fs, repetition rate $f$ = 10 kHz) where the ionization occurs. A static electric field (16 V/cm) guides the ions to a delay-line detector, which records the flight time and impact position for every produced ion. This allows the reconstruction of the 3-dimensional (3D) momenta and the kinetic energies for every detected charged particle. In our experiment, circularly polarized laser pulses with a peak intensity of $3 \times 10^{14}$ W/cm$^2$ are used to ionize $D_2$ dimers. With this intensity and the very low gas density in the reaction region, the overall count rate of photoions in the experiment was below 0.2 ions per laser pulse, resulting in a high-quality coincidence measurement.

**Acknowledgments**

We thank A. R. W. McKellar, A. Stolow, P. Bunker, and T. Pfeifer for fruitful discussions. We acknowledge support from the Joint Centre for Extreme Photonics. Y. M. acknowledges the support from the Deutsche Forschungsgemeinschaft (German Research Foundation) – Grant No. MI 2434/1-1. E. W. is supported by the Strategic Priority Research Program of Chinese Academy of Sciences, Grant No. XDB34020000 and the Alexander von Humboldt Foundation. Financial support from the National Science and Engineering Research Council Discovery Grant No. RGPIN-2020-05858, and from the U.S. Air Force Office of Scientific Research (Grant No. FA9550-16-1-0109) is gratefully acknowledged.

**Author contributions**

Y.M. designed and conducted the experiment; E.W. performed molecular dynamics simulation; Z.D., A.Y.N., and A.S. assisted the experiment; Y. M. and A.S. analyzed the experimental data; Y. M., E.W., and A.S. wrote the manuscript with contributions from all other authors.

**Competing interests**

The authors declare no competing interests.

**Data availability**

The data that support the findings of this study and the raw data of all the figures are available from the corresponding author upon reasonable request.

**Code availability**

The code of the ab-initio molecular dynamics simulation is available from the corresponding author upon reasonable request.

**Additional information**

Supplementary information is available for this paper. Correspondence and requests for materials should be addressed to Y.M.

# Observation and dynamic control of a new pathway of $H_3^+$ formation

# Supplemental Information

Yonghao Mi[1,*], Enliang Wang[2,3], Zack Dube[1], Tian Wang[1], A. Yu. Naumov[1], D. M. Villeneuve[1], P. B. Corkum[1], André Staudte[1*]

[1]*Joint Attosecond Science Laboratory, National Research Council and University of Ottawa, 100 Sussex Drive, Ottawa K1A 0R6, Canada*

[2]*Hefei National Laboratory for Physical Sciences at the Microscale and Department of Modern Physics, University of Science and Technology of China, Hefei, Anhui 230026, China*

[3]*Max Planck Institut für Kernphysik, Saupfercheckweg, 1, 69117 Heidelberg, Germany*

*e-mail: ymi@uottawa.ca; andre.staudte@nrc-cnrc.gc.ca*

## 1. Potential energy curve, orbital character, and charge state

To visualize how the $D_4^+$ dissociates into $D_3^+$ + D channel, we perform an intrinsic reaction coordinate (IRC) calculation from the Franck-Condon (FC) region to the energy minimum of the cation state. Passing the energy minimum, the gradient of the potential energy is too small to perform IRC calculation (due to the flat potential energy surface). We perform the relaxed potential energy scan at large internuclear distance. In this way, we can get the minimum energy reaction path of $D_4^+ \rightarrow D_3^+$ + D. Next, we sort the coordinate according to the D-D bond dissociation and calculate the potential energy.

As shown by the red curve in Fig. S1, the minimum energy path indicates that with increasing internuclear distance between the dissociating $D_2$, the two $D_2$ first come close to each other to form the $D_3$ at the potential energy minimum and dissociate into $D_3^+$ + D channel overcoming the small potential energy barrier of about 0.25 eV.

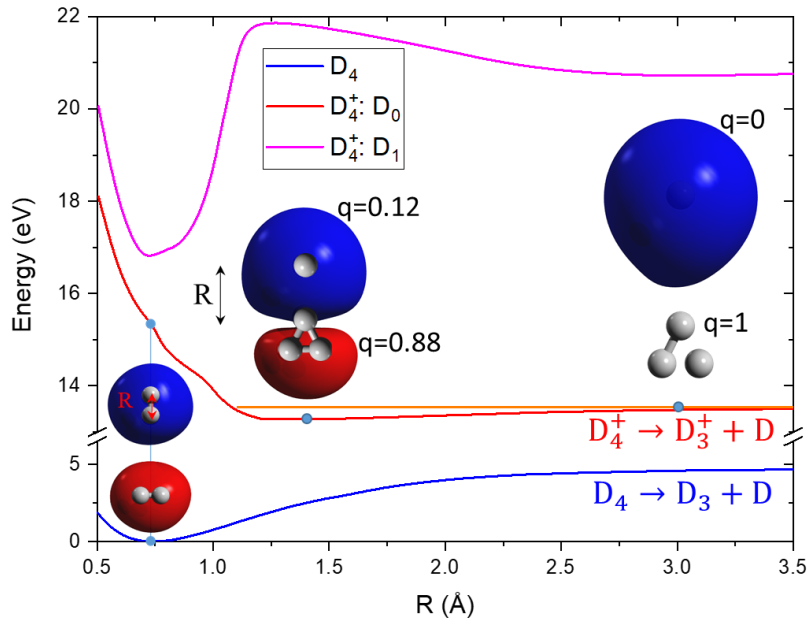

**Fig. S1.** Potential energy curve as a function of the internuclear distance of the dissociating $D_2$. The orange line shows the dissociation limit of the ground cationic state.

The minimum energy reaction coordinate is calculated using the Gaussian package by the second-order Møller–Plesset perturbation theory (MP2) with an aug-cc-pVQZ basis set (*1*). The potential energy is calculated by multistate complete-active-space second-order perturbation theory (MS-CASPT2/aug-cc-pVQZ) using the BAGEL package (*2*). In this calculation, the active space includes two occupied and two unoccupied orbitals and all the orbital electrons are the active ones. In Fig. S1 the blue and the magenta curves show the potential energy of neutral $D_4$ and first excited state $D_4^+$ relative to the energy minima of neutral $D_4$. The wavefunction maps of HOMO orbital of $D_4^+$ show that, with the molecular dissociating, the HOMO orbital will localize on the deuterium atom. Together with the Mulliken charge analysis, it is confirmed that the dissociation limit is $D_3^+ + D$.

As shown in Fig. S1, the energy gap between the ground and the first excited state $D_4^+$ is large enough to avoid crossing between them. In this case, we only consider the ground state $D_4^+$ in the molecular dynamic simulation. The orbital maps and charge states of the DFT calculation of the equilibrium geometries of $D_4$, $D_4^+$, and $D_4^+$ at large internuclear distance are summarized in Tab. S1.

**Tab. S1.** Comparing HOMO wavefunction maps of neutral $D_4$ and $D_4^+$ of Hartree-Fock and density functional theory calculation. The charge state of $D_4^+$ at different internuclear distances is also shown.

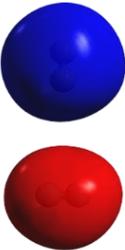

## 2. *Ab initio* molecular dynamics simulation

The *ab initio* molecular dynamics simulation is performed in three steps: (i) the optimized geometry and frequency of the ground state $D_2$ dimer are calculated using the MP2/aug-cc-pVQZ method. In the second step (ii) we are sampling the zero-point vibration of the neutral $D_2$ dimer by the Wigner distribution which was calculated using the Newton-X package (*3-4*). Here, we initialize the $D_2$ dimer with the T-shape geometry. Due to the low potential energy barriers, the full potential energy surface within the Franck-Condon region can be sampled even using only one initial equilibrium geometry. As shown in Fig. S2, the sampled relative molecular axis angles include all the possible geometries of the equilibrium $D_2$ dimer (*5*). Finally, in the third step (iii) we perform the molecular dynamics simulation in the extended Lagrangian molecular dynamics scheme, adopting the so-called atom-centered density matrix propagation (ADMP) method using the density-functional theory method at ωb97xd/aug-cc-pVDZ level. We assume a vertical ionization from the neutral to the cation state. Thereby the geometry and velocity sampled from step (ii) are used as the initial configuration of the simulation. The fictitious electron mass is 0.1 amu, and the simulation time step was 0.5 fs. To obtain adiabatic control, we run the ADMP simulation with the converged self-consistent field at each step which yields an equivalent result to the Born–Oppenheimer molecular dynamics.

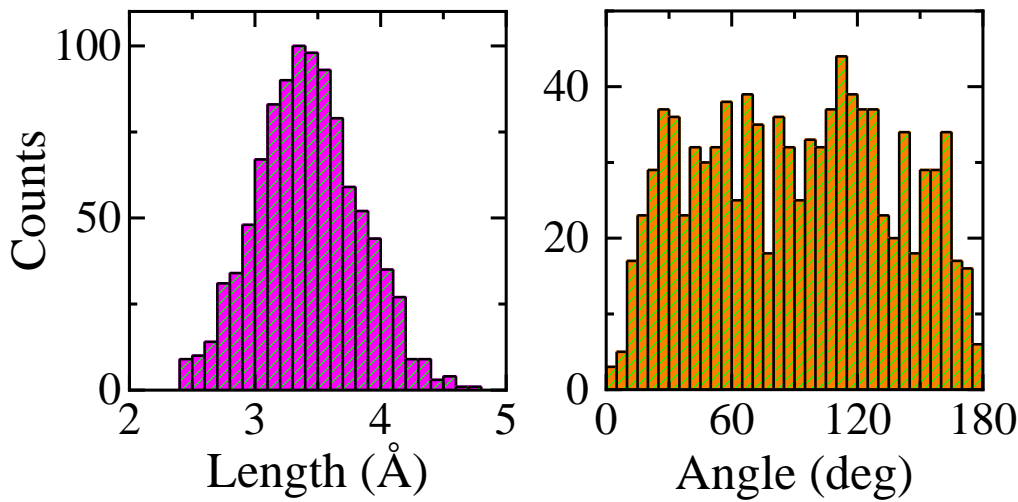

**Fig. S2.** The sampled center-of-mass distances and relative molecular axis angles of the $D_2$ dimer.

### 3. Anharmonic effect in the sampling

The Wigner distribution is based on harmonic oscillator functions. Due to the flat potential energy surface of $D_2$-$D_2$ the anharmonic effect cannot be ignored. We calculated the potential energy curve as a function of the center of mass (COM) distance between $D_2$ and $D_2$ by CCSD(T)/aug-cc-PVQZ, as shown by the blue curve in Fig. S3 (a). We put together the sampled COM distance distribution, as shown by the magenta bars in Fig. S3(a). Obviously, all the possible regions of the ground potential energy surface are sampled, however, the sampled initial geometries overestimate the smaller internuclear distance which is outside the inner classical turning point and of about 7% of the total samples. The molecular dynamics simulation shows that the smaller initial internuclear distance also results in the same asymptotic channel: $D_4^+ \rightarrow D_3^+ +$ D. As shown in Fig. S3 (b), we compare the COM distance between $D_3^+$ and D as a function of time with 2.5 Å and 3.4 Å initial COM distance and both without initial internal energy. The smaller initial internuclear distance results in a short formation time.

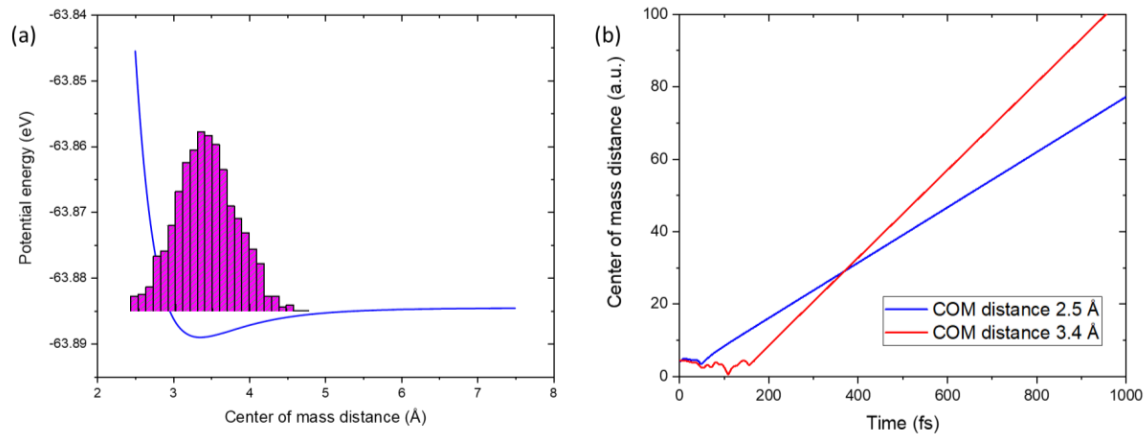

**Fig. S3.** (a) Potential energy curve (blue curve) as a function of the COM distance between $D_2$ and $D_2$ by CCSD(T)/aug-cc-PVQZ and sampled COM distance distribution (magenta bars). (b) Compering the molecular dynamics simulation of different initial COM distances. The 2.5 Å and 3.4 Å correspond to the smallest COM distance of the sample and the equilibrium geometry, respectively.

## 4. Simulation of $H_2$ dimer

To compare the $H_3^+$ formation branching ratio from $H_2$ dimer, we perform molecular dynamics simulation of $H_4^+$ by means of the same method of $D_4^+$ simulation. The comparison of time-dependent center-of-mass distance for a T-shaped and a parallel-aligned $D_2$ and $H_2$ dimer are shown in Fig. S4. To estimate the branching ratio of $H_3^+$ formation, we simulate 593 trajectories of $H_4^+$ and 493 of them end up with $H_3^+ + H$ channel (see Fig. S5). A branching ratio of 83.1% is obtained which is very close to that of the $D_3^+$ (89.2%).

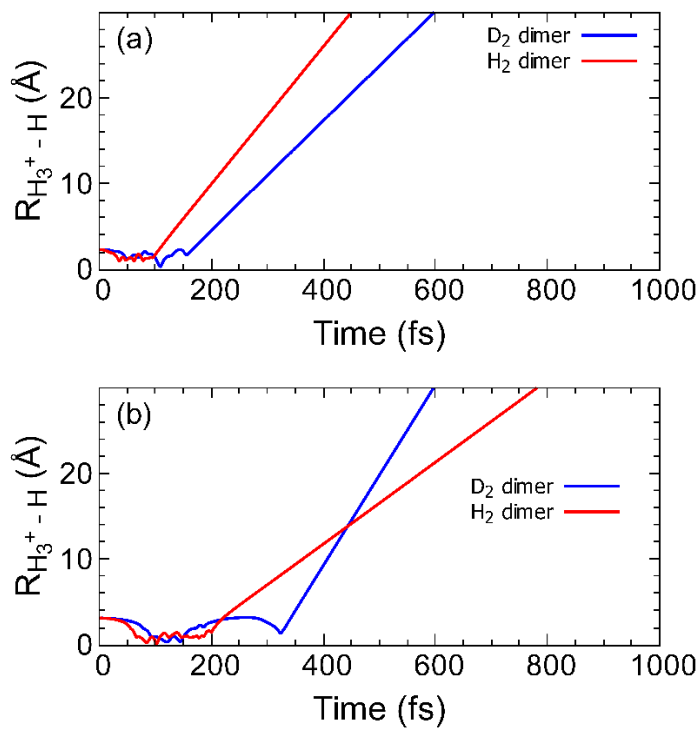

**Fig. S4**. Simulation results of the time-dependent center-of-mass distance for a T-shaped (a) and a parallel-aligned (b) dimer.

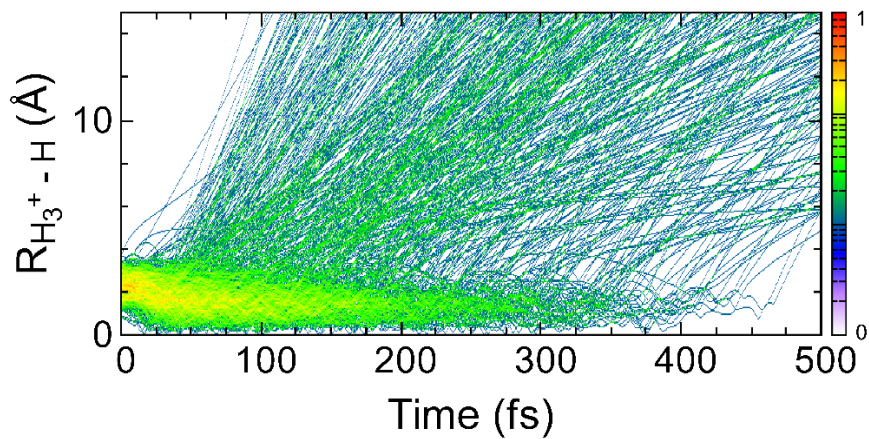

**Fig. S5**. Simulated center-of-mass distance of $H_3^+ + H$ channel for 493 trajectories.

## 5. Kinetic energy release (KER) distribution

We show the kinetic energy release (KER) spectrum for the following three channels in Fig. S6: $D + D_3^+$, $D_2^+ + D_2^+$ and $D^+ + D_3^+$. The KER of the double ionization channel $D_2^+ + D_2^+$ (black curve) is centered at 3.7 eV. The other double ionization channel $D^+ + D_3^+$ (red curve), however, leads to a higher KER (4.4 eV). This difference indicates that $D^+ + D_3^+$ occurs at a shorter intermolecular distance. Most $D_3^+$ are produced from single ionization $D_3^+ + D$ (blue curve), with a KER centered at 0 eV. This indicates that the $D_3^+$ is created in a statistical dissociation (6).

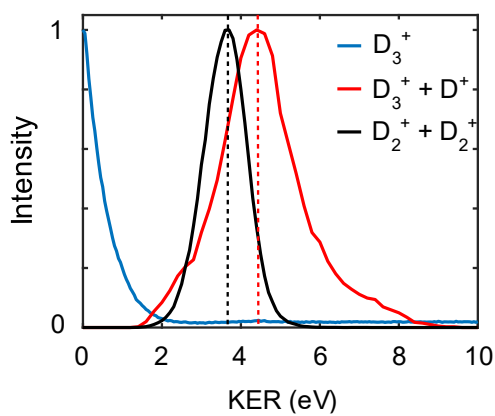

**Fig. S6.** Kinetic energy release (KER) distribution for different channels

## 6. Suppressed yield of the $D_3^+ + D$ channel

The suppressed yield of $D_3^+ + D$ for delays >200fs should reappear as a rise of a different channel, with the double ionization channel $D_3^+ + D^+$ being the natural candidate for it. In fact, in Fig.3c of the manuscript one can see that rise. However, the amount of detected $D_3^+ + D^+$ is two orders of magnitude lower than the drop in the $D_3^+ + D$ channel. Such a discrepancy cannot be explained by the lower double hit detection efficiency (as compared to the single hit detection efficiency), but must be attributed to the low intensity of the probe pulse which apparently only ionizes at most a few percent of any available atomic hydrogen.

Consequently, we should not expect the two other possible double ionization channels to contribute either:

a) $(D_2)_2^+ \rightarrow D_2^+ + D_2^+$: As shown in Fig.3b) in the manuscript, this channel does not only have an insufficient number of events, but actually drops at the same time as the $D_3^+ + D$ pathway. Since $D_2$ has a ~1.5eV higher ionization potential than that of atomic deuterium, we expect an even smaller yield of $D_2^+$ from the probe pulse.

b) $(D_2)_2^+ \rightarrow D_2^+ + D^+ + D$: Here, $D_2^+$ is being produced with another $D^+$ and D. While we cannot measure the neutral D, we can assume that the main part of the kinetic energy release is shared among the charged particles. We analysed our data for a weakly correlated $D_2^+ + D^+$ channel, and found only an enhancement for delays < 100fs, i.e., in the overlap region of both pulses (see Fig. S7).

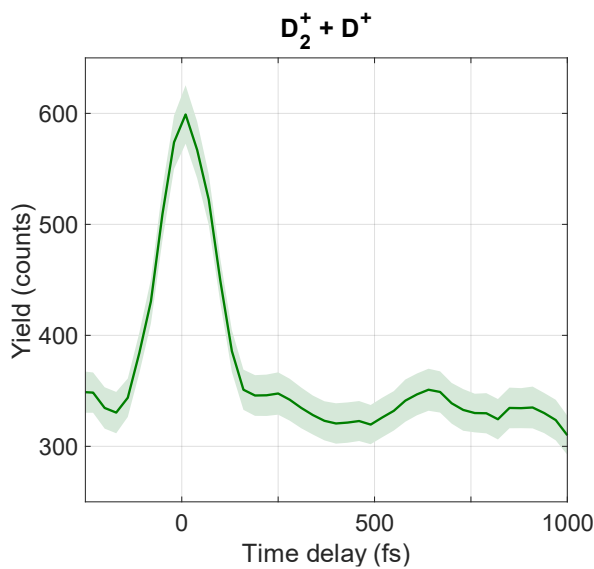

**Fig. S7**. Measured time-dependent yield for the $D_2^+ + D^+$ channel

Hence, only three single ionization pathways remain to explain the dip in $D_3^+$ production:

c) $(D_2)_2^+ \rightarrow D_2^+ + D_2$: Unfortunately, the $D_2^+$ from this pathway coincides with a huge $D_2^+$ peak from the monomer single ionization. With a bond energy of only 3 meV a dissociation of the singly charged dimer into two monomers will not produce significant momentum in the $D_2^+$ to be recognizable below the monomer velocity spread. Thus, we cannot quantify this pathway.

d) $(D_2)_2^+ \rightarrow D_2^+ + D + D$: This pathway assumes that the association of the hydrogen atom with the molecular ion is being interrupted by perturbing (Stark-shifting) the potential energy surface of the $D_4^+$. The only charged particle in this channel, the $D_2^+$ is also competing with the dominant single ionization of the monomer. Therefore, it is also challenging to isolate this pathway. However, neither the deuterium molecule nor the atom is very polarizable, and therefore are not susceptible to the probe pulse's electric field. On the other hand, the $D_2^+$ molecule in the dimer is very susceptible, as we can see in the simulation.

e) $(D_2)_2^+ \rightarrow D^+ + D + D_2$: While the probe pulse is not intense enough to efficiently multiphoton-ionize either $D_2$ or $D$, it is intense enough to cause few-photon transitions in the molecular ion. In $D_2^+$ the $1s\sigma_g$ ground state and the $2p\sigma_u$ dissociative, first excited electronic state can be coupled by an odd number of photons, which allows the molecular ion to dissociate in the presence of the laser field. This process has been studied since the early 1990's and is called *bond softening* and *above threshold dissociation* (7). More recently this process has been examined from the perspective of light-induced conical intersections (8). Through bond softening $H_2^+$ fragments into a proton and a hydrogen atom, with the kinetic energy corresponding to the number of photons involved. The fragment distribution depends on the polarization of the light field, that is, in circular polarization the proton momenta are distributed in rings in the polarization plane, whereas in linearly polarized light the proton momenta are aligned with the polarization axis, forming peaks that correspond to the net-number of absorbed photons. In our data we found clear evidence that the probe pulse is very efficient in dissociating $D_2^+$ molecules. Since the molecular ion is a part of the $(D_2)_2^+$ it can be bond softened, while the dimer is being formed. Unfortunately, the bond softening of $D_2^+$ from the dominant monomer cannot be distinguished from the dimer pathway. However, in a direct comparison of the low-energy ($|pz| < 6$ a.u.), polarization-aligned $D^+$ (as shown by the red dashed rectangular in Fig. S8 (a)) yield with the $D_3^+$ yield we find a slight enhancement of the $D^+$ production for positive delays, where the $D_3^+$ yield drops below its asymptotic value.

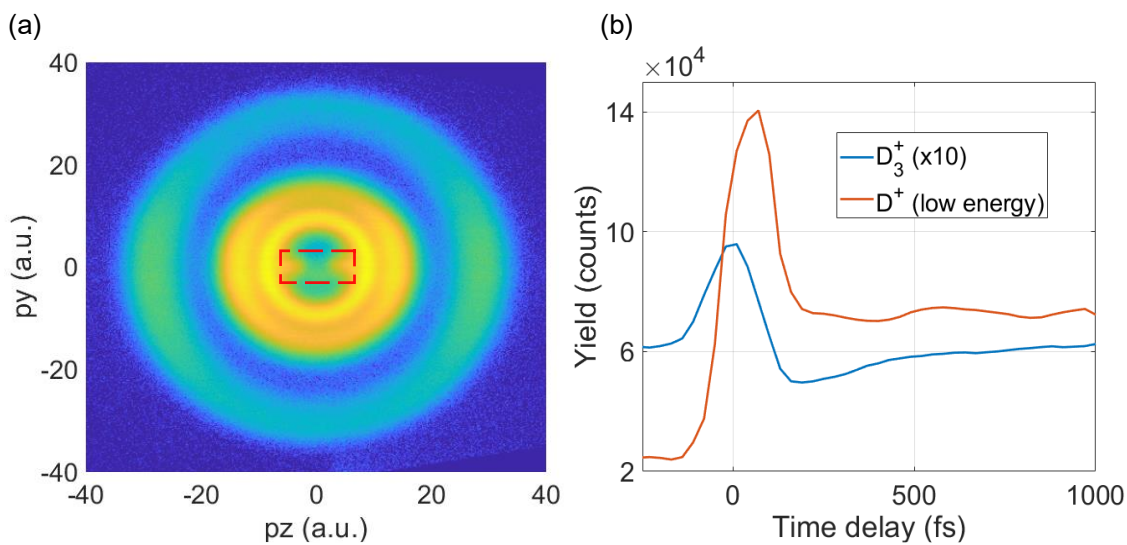

**Fig. S8**. Momentum distribution of $D^+$ and the time-dependent yields of the low-energy $D^+$ (red curve) and $D_3^+$ (blue curve, actual yield $\times 10$).

**Movie S1.**

The formation process of $D_3^+$ for a parallel-aligned $D_2$ dimer

**Movie S2.**

The formation process of $D_3^+$ for a T-shape $D_2$ dimer